\documentclass[aps,prl,twocolumn,superscriptaddress,showpacs]{revtex4}
\usepackage{graphicx}

\newcommand{\grad}{\mbox{\(\mathsurround=0pt{}^\circ\)}}

\begin{document}

\title{Phase shift experiments identifying Kramers doublets in
a chaotic superconducting microwave billiard of threefold symmetry}

\author{C.~Dembowski}
\affiliation{Institut f\"ur Kernphysik, Technische Universit\"at Darmstadt,
D-64289 Darmstadt, Germany}
\author{B.~Dietz}
\affiliation{Institut f\"ur Kernphysik, Technische Universit\"at Darmstadt,
D-64289 Darmstadt, Germany}
\author{H.-D.~Gr{\"a}f}
\affiliation{Institut f\"ur Kernphysik, Technische Universit\"at Darmstadt,
D-64289 Darmstadt, Germany}
\author{A.~Heine}
\affiliation{Institut f\"ur Kernphysik, Technische Universit\"at Darmstadt,
D-64289 Darmstadt, Germany}
\author{F.~Leyvraz}
\affiliation{Centro de Ciencias F{\'\i}sicas, UNAM, Mexico}
\author{M.~Miski-Oglu}
\affiliation{Institut f\"ur Kernphysik, Technische Universit\"at Darmstadt,
D-64289 Darmstadt, Germany}
\author{A.~Richter}
\affiliation{Institut f\"ur Kernphysik, Technische Universit\"at Darmstadt,
D-64289 Darmstadt, Germany}
\author{T.~H.~Seligman}
\affiliation{Centro de Ciencias F{\'\i}sicas, UNAM, Mexico}

\date{\today}

\begin{abstract}
The spectral properties of a two-dimensional microwave billiard showing
threefold symmetry have been studied with a new experimental
technique. This method is based on the behavior of the eigenmodes under
variation of a phase shift between two input channels, which strongly
depends on the symmetries of the eigenfunctions.
Thereby a complete set of $108$ Kramers doublets has been identified
by a simple and purely experimental method. This set clearly shows Gaussian
unitary ensemble statistics, although the system is time-reversal
invariant.
\end{abstract}

\pacs{05.45.Mt, 11.30.Er}

\maketitle

In a recent experiment with a superconducting microwave resonator,
which possesses threefold symmetry ($C_{3}$ symmetry),
first signatures for spectral properties according to
Gaussian unitary ensemble (GUE) statistics have been observed for
a time-reversal invariant system \cite{dreieck1}.
Discrepancies between experiment and theory remained, although these
results confirmed the theoretical prediction \cite{lss}, that time-reversal
invariant systems may show properties,
characteristic of violation of time-reveral invariance (TRI).
A comparison with numerical results allowed to understand the
discrepancies, which
were due to a less than perfect identification of Kramers doublets
that are supposed to display GUE statistics. 
The only criterion used to identify these doublets, which are split due 
to $C_{3}$  symmetry breaking perturbations, was the spacing between
consecutive resonances. The results of \cite{stoeckmann_dreieck} were even
worse and demanded for an explanation by ambitious numerical studies. 
In the present paper
we report on a new experimental technique, which identifies the
doublets properly and uniquely, and agreement between experiment
and theory is obtained within statistical uncertainty.

\begin{figure}[hb]
\centerline{\includegraphics[width=8.7cm]{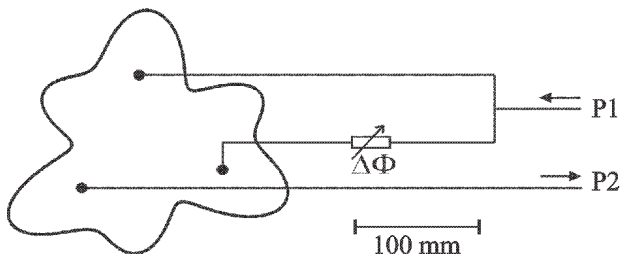}}
\vspace{0.5cm}
\caption{Schematic view of the experimental setup. The chaotic microwave
cavity (left side) possesses threefold symmetry. The microwaves
provided by port 1 (P1) of the network analyzer are split into two
signals before they are coupled
into the cavity by two different channels. The phase shift
$\Delta\Phi$ between these two channels can be varied by a phase shifter.
Power is coupled out by a third antenna and led back to the network
analyzer, where the transmission to port 2 (P2) is measured.}
\label{setup}
\end{figure}

Billiards are model systems, which allow to study the properties
of classically chaotic dynamics: a point-like
particle moving without losses inside a closed boundary, where
it is reflected specularly at impact,
possesses non-integrable equations of motion, if the boundary
is suitably shaped (see, e.g., \cite{berry81}). These systems have
a quantum mechanical analogue, where the
chaotic behavior manifests itself in the properties of the
eigenfunctions and spectra \cite{bohigas}.
It is common knowledge that the statistical fluctuations
of the eigenvalues of a quantum system with a
classically chaotic counterpart can be described by
random matrix theory \cite{bohgiaschm,mehtaguhr}.
For chaotic quantum systems one usually finds fluctuations
similar to those of the Gaussian orthogonal ensemble (GOE),
which reflects TRI. It has been recognized,
however, that certain systems with TRI show spectral fluctuations
which follow the GUE and thus mimic a violation of TRI \cite{lss}.
A simple example for a time-reversal invariant
system with GUE statistics is a two-dimensional chaotic billiard, the
highest symmetry of which is $C_{3}$. Such a billiard has been used in
\cite{dreieck1} and in the present work (see fig.~\ref{setup}).
As was pointed out in \cite{dreieck1,lss}
the Hamiltonian of this triangular billiard possesses three classes of
eigenfunctions $\Psi^{(l)}$ with different symmetry properties ($l=-1,0,+1$).
A rotation $R$ of $120\grad$ transforms these eigenfunctions according to
$R\Psi^{(l)}={\rm exp}(2\pi i l /3)\Psi^{(l)}.$
Two of the classes of wave functions are degenerate due to TRI and
one finds a spectrum composed of singlet ($l=0$) and doublet ($l=\pm 1$)
modes. The doublets were predicted to show GUE statistics, while the
singlets should follow GOE statistics. In a complex representation
time-reversal will transform one eigenfunction of a doublet into its
complex conjugate and thus interchanges the two eigenfunctions, i.e.
$\Psi^{(-1)}=T\Psi^{(+1)}=(\Psi^{(+1)})^{*}$.
This is an example for Kramers degeneracy: 
whenever TRI holds, i.e. $[H,T]=0$, the time-reversed quantity
$T\Psi$ of an eigenfunction $\Psi$ is also an eigenfunction
of $H$. Thus, if $\Psi \neq T\Psi$, TRI forces a
degeneracy (see, e.g., \cite{kramers}).
This is known as Kramers theorem and its consequences play an important
role in studies of systems with half-integer spin
\cite{beispiele}.

Two-dimensional quantum billiards can be studied experimentally with
the help of macroscopic analog systems \cite{stoeckmannbuch,richter}:
in flat cylindric resonators only transverse magnetic modes exist
below a critical frequency.
These modes have electric field vectors perpendicular to the bottom
and lid of the resonator, while the magnetic field lines lie parallel
to them. The electric field
$\vec{E}_{n}(\vec{r})=\phi_{n}(x,y)\vec{e}_{z}$
as well as the magnetic field is fully described by the solutions
$\phi_{n}(x,y)$ of the scalar Helmholtz equation
\begin{equation}
(\Delta+k^{2}_{n})\phi_{n}=0
\end{equation}
with Dirichlet boundary conditions.
There is a complete analogy
between the two-dimensional Helmholtz equation for a flat microwave
cavity and the Schr{\"o}dinger equation describing a quantum billiard
of the same shape.
The wave numbers $k_{n}$ of the electromagnetic resonator correspond to
the energy eigenvalues $\varepsilon_{n}$ of the Hamiltonian of the
quantum billiard, i.e. $k_{n}^{2}\propto \varepsilon_{n}$.

\begin{figure}
\centerline{\includegraphics[width=8.7cm]{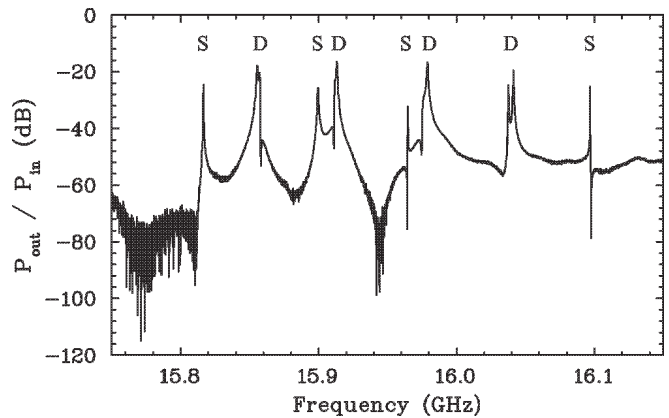}}
\vspace{0.5cm}
\caption{Part of a transmission spectrum measured at $T=4.2~K$.
The maxima of the resonance line mark the resonance frequencies.
One clearly observes a signal-to-noise ratio of about four
orders of magnitude (40~dB). Singlets and doublets are
marked by S and D, resp.} 
\label{bildspektrum}
\end{figure}

The usual way to determine the eigenfrequencies of a microwave cavity
$f_{n}=k_{n}c/(2\pi)$, where $c$ denotes the speed of light, is to
measure the power transmitted from one antenna emitting microwaves
into the cavity to another antenna receiving the microwaves.
As a result one gets a resonance line (see fig.~\ref{bildspektrum}),
where the positions of the maxima
correspond to the eigenfrequencies $f_{n}$, while the
widths of the resonances $\Delta f_{n}$ reflect the damping,
characterized by the quality factor $Q=f_{n}/\Delta f_{n}$. The
widths of the resonances define the experimental
limit of resolution. Thus, a superconducting cavity of high $Q$
is a prerequisite for resolving small splittings of degenerate
eigenvalues, which stem from mechanical imperfections, and
for identifying Kramers doublets properly.
In our experiment a resonator made from lead plated
copper was used. The measurements were carried out in a cryostat with
liquid helium at a temperature of 4.2~K. The resonator becomes
superconducting at temperatures below approximately 7.2~K and
possesses a quality factor of the order of $10^4$.
The microwave signals are processed by a HP-8510C
network analyzer, which also controls the signal generator providing
the microwaves. This technique has been used in several experiments
\cite{stoeckmannbuch,richter}, including the
first measurements with the cavity under consideration \cite{dreieck1},
and now was modified to improve the results in the
following way (see fig.~\ref{setup}).

An {\it additional} antenna is used for feeding microwave power
into the resonator. Thus, power is
coupled into the system through two different channels at the
same time, while the output power is coupled out through a third
channel. The positions of the three antennas were chosen in a way,
which preserves the threefold symmetry of the billiard. A Midwest Microwave
PWD-2533-02-SMA-79 power divider splits the microwave power
provided by the signal generator into two signals.
These two input signals have a certain phase shift 
\begin{equation}
\Delta\Phi_{total}=\Delta\Phi+\Delta\Phi_{0},
\end{equation}
where $\Delta\Phi$ can be varied by an ARRA~9428A
phase shifter, while $\Delta\Phi_{0}$ is an offset, which is constant for
a given frequency and is due to the different lengths of the two signal paths.
The phase shifter and the power divider operate up
to a frequency of 18~GHz. For frequencies below 6~GHz the
maximum achievable $\Delta\Phi$ is less than $360\grad$.
For the frequency range from 8~GHz up to 18~GHz 50 different transmission
spectra were measured, covering phase shifts from $\Delta\Phi\approx~0\grad$
up to $\Delta\Phi\approx~350\grad$ in steps of $7\grad$.
Data points were taken in steps of 100~kHz, i.e. a resolution
of the order of $10^{-5}$, which has to be compared with $1/Q \approx 10^{-4}$.
Furthermore, conventional transmission spectra -- as described above --
have been recorded in the frequency range from 0 to 18~GHz 
with a spectral resolution of 50~kHz (see fig.~\ref{bildspektrum}). This was
done for two different antenna combinations, with a fourth antenna
not used for the phase shift experiment. These transmission spectra
served to identify the resonance frequencies, while the results of
the phase shift measurements were utilized to identify the eigenmodes as
singlets or doublets.

\begin{figure}
\centerline{\includegraphics[width=8.7cm]{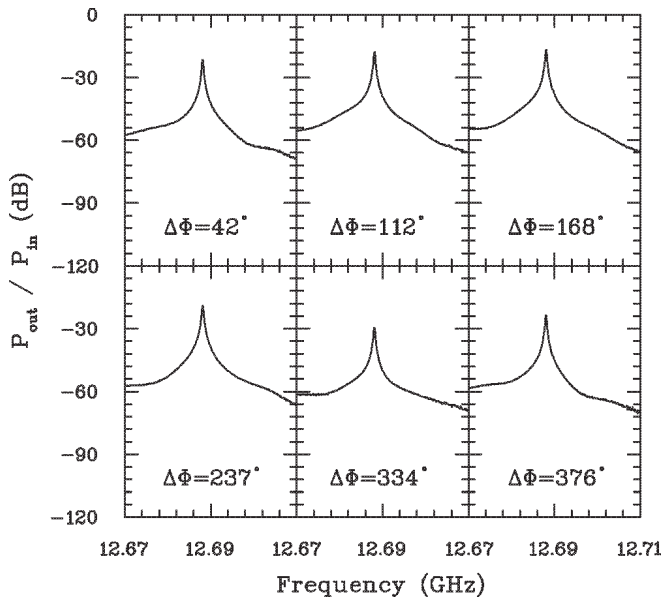}}
\vspace{0.5cm}
\caption{Resonance line for a typical singlet.
The {\it shape} of the line depends only weakly on the phase shift varied.
Nevertheless, there is a change in amplitude of more than one order of
magnitude, i.e. a difference of more than 10~dB in the logarithmic scale.}
\label{singlet}
\end{figure}

As we had conjectured, singlets and doublets typically show
a very different behavior, when the phase shifter setting $\Delta\Phi$
is varied.
For a typical singlet mode (fig.~\ref{singlet})
the {\it shape} of the resonance line does not change,
although there is a change in the amplitude: 
the transmitted power at resonance depends on $\Delta\Phi$
and shows a minimum and a maximum, corresponding to destructive or
constructive interferences. Because of the special symmetries
of the antennas and of the singlet wave functions these should occur
at a distance of $180\grad$. However, the symmetries are not perfect
and thus the minimum amplitude is non zero, but for
some singlets the intensities at the
maximum and the minimum differ by two orders of magnitude. Furthermore
minimum and maximum do not always have an exact phase difference
of $180\grad$ as expected, but $179\grad \pm 33\grad$. There is a
correlation between the deviation from $180\grad$ and the minimum
intensity. Large shifts only appear, when the minimum and the maximum
intensity differ only marginally.
The shape of the resonance, however, does not change due to those effects.

\begin{figure}
\centerline{\includegraphics[width=8.7cm]{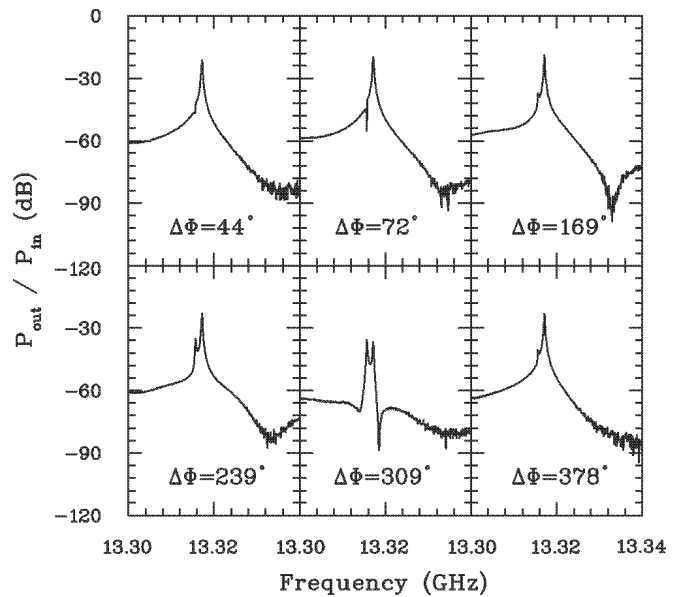}}
\vspace{0.5cm}
\caption{Resonance line for a typical doublet.
The shape of the line strongly depends on the phase shift varied.
Note that this is a semi-logarithmic plot, which
compresses the changes in amplitude.}
\label{doublet}
\end{figure}

In contrast to the singlet levels, typical doublets show a strong variation
of the shape of the resonance curve (fig.~\ref{doublet}).
The relative and the absolute height of each of the two peaks
within a doublet depends on the phase
difference and sometimes an almost complete vanishing of one of the
peaks is achieved. In some cases a doublet can only be identified
for certain values of the phase shift. However, the behavior of the
doublet amplitudes depending on $\Delta\Phi$ is much harder to understand,
as one has to deal with two eigenmodes being excited simultaneously with
different amplitudes, the ratio of which depends on the properties of
the two individual states involved and is uncontrollable in our setup.
Nevertheless, a strong influence of the induced phase shift on the
shape of a resonance line composed of a doublet is obvious
(see fig.~\ref{doublet}), which provides an important
additional criterion for distinguishing doublets from singlets. 

\begin{figure}
\centerline{\includegraphics[width=8.7cm]{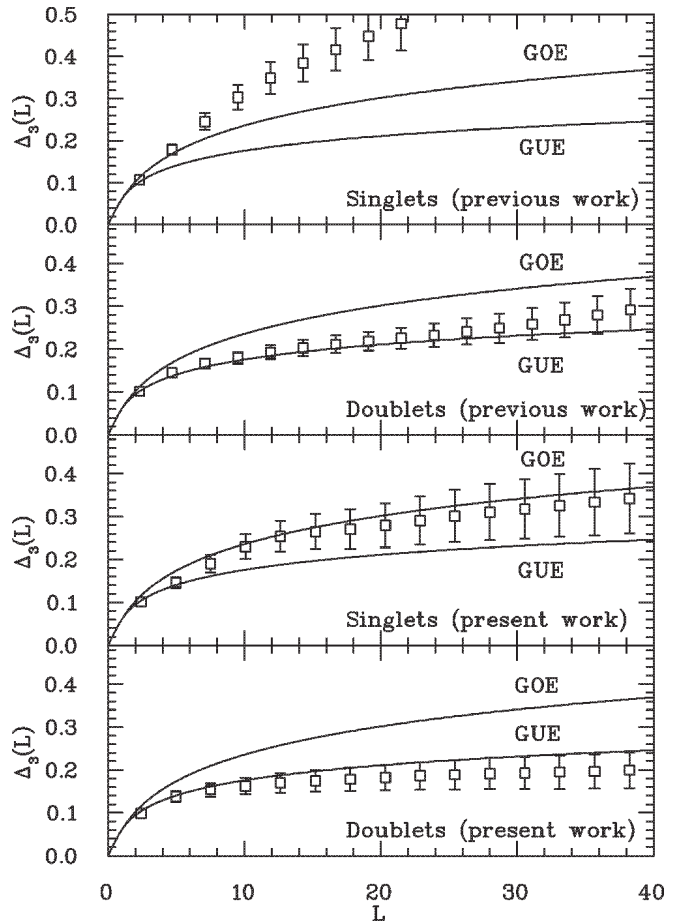}}
\vspace{0.5cm}
\caption{Long range correlations for the singlets and doublets:
in the upper half the Dyson-Mehta statistics obtained in the framework
of our previous work is shown. As one can see there
are distinct deviations from GOE for the singlets. Also there is no
saturation behavior for singlets and doublets.
In the lower part the Dyson-Mehta statistics for
the results of the present work are plotted. The statistics show for both
singlets and doublets saturation, although the singlets still deviate
slightly from GOE for small interval lengths $L$.}
\label{delta3}
\end{figure}

With the help of the method described above, 102 singlets and 108
doublets have been identified unambigously in the frequency range
up to 18~GHz. The sets of eigenfrequencies {$f_i$} were unfolded
using Weyl's law \cite{balteshilf}:
\begin{equation}
x_{i}:=N_{Weyl}(f_{i})=v_{1}f_{i}^{2}+v_{2}f_{i}+v_{3}
\label{weylformula}
\end{equation}
This procedure leads to spectra {$x_i$} with
normalized mean level spacings
of 0.988 and 1.006 for singlets and doublets, respectively.
As the number of levels in the two spectra is only about 100, we chose to analyze
the long-range correlations of the spectra.
Evaluating the Dyson-Mehta $(\Delta_{3})$ statistics
\cite{berry_delta3} yields perfect GUE behavior for the doublets
(see fig.~\ref{delta3}), including saturation \cite{berry_delta3,seligman84},
i.e. $\Delta_{3}(L)$ stops increasing for $L>L_{max}$. The
value of $L_{max} \approx 10$ is consistent with an
estimate of the length of the shortest peridic orbit in the billiard.
All this indicates that the corresponding
spectrum is indeed {\it complete}. Otherwise deviations would occur,
as observed in our previous work \cite{dreieck1}.
For the singlets we still observe deviations from GOE for small
spacings (see fig.~\ref{delta3}).
As there are fewer singlets than doublets, this has to be due to missing
levels. A check with a newly developed method based upon the assumptions
that resonances below a certain critical strength are not detected and
that the positions of the missing levels are distributed randomly
indeed indicates that we lost about 5 percent of
the singlets, while we have all of the doublets \cite{mitchell}.
It is most likely that some of the singlet modes cannot be detected
with our antennas, because their positions show the same symmetry
properties ($C_{3}$) as the singlet wave functions.
We cannot circumvent this problem without destroying the $C_{3}$ symmetry,
e.g. when inducing a perturbing body.
Thus, if one antenna lies on a nodal line, all antennas do so at 
frequencies corresponding to singlets, while this
is not the case for the doublets. These considerations are also supported
by results we obtain for the Weyl coefficients $v_{1}$, $v_{2}$, and
$v_{3}$, which were determined by a fit of eq.~(\ref{weylformula})
to the experimental data. The coefficients $v_{1}$ and $v_{2}$ are
connected to area $A$ and perimeter $U$, resp.,  of the billiard.
We found
$A~=~0.03~m^{2}$ and $U~=~1.16~m$ for the singlet spectrum and
$A~=~0.03~m^{2}$ and $U~=~0.81~m$ for the doublets. Therefore, the Weyl
parameters of the doublet spectrum are much closer to the
the physical dimensions of the cavity, which are
approximately $A~=~0.03~m^{2}$ and $U~=~0.82~m$ at room temperature. Note,
that for Dirichlet boundary conditions a larger perimeter $U$ corresponds
to a smaller number of eigenfrequencies \cite{balteshilf}.

We can be sure, however, not to have spurious singlet levels.
This was the case in~\cite{dreieck1}, where we did not have a reliable
method of separating the subspectra, which caused misassignments,
i.e. lost {\it and} extra
eigenvalues in the resulting spectra of 196 singlets and 213 doublets
up to 25~GHz. Thus, these spectra showed deviations from GOE- and
GUE-like statistics, respectively \cite{dreieck1}.
Also there was no saturation behavior of the Dyson-Mehta
statistics (see fig.~\ref{delta3}).
Now saturation is obtained both for singlets and doublets.
Although the modified setup restricts us to a smaller frequency range,
it nevertheless allows a much more reliable identification and classification
of states as singlets or doublets.

To conclude, the spectrum of a microwave cavity with $C_{3}$ symmetry shows
degenerate and non-degenerate modes. Under variation of a phase shift
between two feeding antennas these classes
of states behave differently. Therefore, the experimental technique
presented allows us for the first time to classify
singlets and doublets as such. For the doublet spectrum fluctuation
properties of the eigenvalues perfectly matching GUE statistics 
have been observed for a system with TRI -- with 108 Kramers doublets only.
This constitutes a clear improvement upon earlier results,
where additional numerical
information was needed to fully identify doublets \cite{dreieck1}
or to explain the effects of level splitting and level
loss \cite{stoeckmann_dreieck}.

We would like to thank H.~L.~Harney and G.~E.~Mitchell for many fruitful discussions.
We acknowledge the DFG for supporting this work under contract no. 
Ri~242/16-3. C.D., B.D., and A.H. thank CONACyT for support
during the workshop on {\it Chaos in few and many body problems} at
CIC. B.D. thanks the HMWK for support within the HWP.
F.L. and T.H.S. acknowledge support by DGAPA.


\begin{thebibliography}{99}

\bibitem{dreieck1}C.~Dembowski, H.-D.~Gr{\"a}f, A.~Heine, H.~Rehfeld,
A.~Richter, and C.~Schmit, Phys. Rev. E {\bf 62} (2000) R4516.

\bibitem{lss} F.~Leyvraz, C.~Schmit, and T.~H.~Seligman, J. Phys. A {\bf 29}
(1996) L575.

\bibitem{stoeckmann_dreieck}R. Sch{\"a}fer, M.~Barth, F.~Leyvraz,
M.~M{\"u}ller, T.~H.~Seligman, and H.-J.~St{\"o}ckmann,
Phys. Rev. E {\bf 66} (2002) 016202. 

\bibitem{berry81}M.~V.~Berry, Eur. J. Phys. {\bf 2} (1981) 91.

\bibitem{bohigas} O.~Bohigas, in
{\it Chaos and Quantum Physics}, edited by M. J. Gianonni, A. Voros, and J.
Zinn-Justin (Elsevier, Amsterdam, 1991).

\bibitem{bohgiaschm} O.~Bohigas, M.~J.~Giannoni, and C.~Schmit, Phys. Rev.
Lett. {\bf 52} (1984) 1; A.~V.~Andreev, O.~Agam, B.~D.~Simons, and
B.~L.~Altshuler, Phys. Rev. Lett. {\bf 76} (1995) 3947; F.~Leyvraz and
T.~H.~Seligman, in {\it Proceedings of the Fourth Wigner Symposium},
edited by N.~Atakishiev, T.~H.~Seligman, and K.~B.~Wolf, pp. 429 (World
Scientific, Singapore, 1996).

\bibitem{mehtaguhr} M.~L.~Mehta, {\it Random Matrices and the
Statistical Theory of Energy Levels}, 2nd. ed. (Academic Press, San Diego,
1991); T.~Guhr, A.~M{\"u}ller-Groeling, and H.~A.~Weidenm{\"u}ller,
Phys. Rep. {\bf 299} (1998) 189.

\bibitem{kramers}
R.~ Scharf, B.~Dietz, M.~Ku{\'s}, F.~Haake, and M.~V.~Berry,
Europhys. Lett. {\bf 5} (1988) 383.

\bibitem{beispiele}
X.~Li and J.~Dudek, Phys. Rev. C {\bf 49} (1994) R1250;
G.~Usaj and H.~U.~Baranger, Phys. Rev. B {\bf 63} (2001) 184418;
C.~Cascales, C.~Zaldo, U.~Caldi{\~n}o, J.~Garc{\'\i}a Sol{\'e},
and Z.~D.~Luo, J. Phys.: Condens. Matter {\bf 13} (2001) 8071;
S.~Jandl, P.~Richard, M.~Poirier, V.~Nekvasil, A.~A.~Nugroho, A.~A.~Menovsky,
D.~I.~Zhigunov, S.~N.~Barilo, and S.~V.~Shirayaev,
Phys. Rev. B {\bf 61} (2000) 12882.

\bibitem{stoeckmannbuch} H.-J.~St\"ockmann, {\it Quantum Chaos:
An Introduction} (Cambridge University Press, Cambridge, 1999).

\bibitem{richter} A.~Richter,
in {\it Emerging Applications of Number Theory}, The IMA Volumes in
Mathematics and its Applications, Vol. {\bf 109}, edited by D.~A.~Hejhal,
J.~Friedman, M.~C.~Gutzwiller, and A.~M.~Odlyzko, pp. 479
(Springer, New York, 1999).

\bibitem{balteshilf}H.~P.~Baltes and E.~R.~Hilf, {\it Spectra of Finite
Systems} (Bibliographisches Institut, Mannheim, 1976).

\bibitem{berry_delta3} M.~V.~Berry, Proc. R. Soc. London, Ser. A {\bf 400}
(1985) 229.

\bibitem{seligman84} T.~H.~Seligman, J.~J.~M.~Verbaarschot, and
M.~R.~Zirnbauer, Phys. Rev. Lett. {\bf 53} (1984) 215.  

\bibitem{mitchell}U.~Agvaanluvsan and G.~E.~Mitchell, private
communication (2002).


\end{thebibliography}
\end{document}